----------------------------------------------------------------------
\documentstyle[12pt,epsf]{article}
\begin{document}
\noindent {\it Submitted to Phys.\ Lett.\ B} \hfill DOE/ER/40762--031

\hfill U. of MD PP \#94--101

\vspace{32.4pt}
\centerline{\bf Disoriented Chiral Condensates and }
\centerline{\bf Quantum Mechanical Isospin Correlation }
\vspace{14.4pt}
\centerline{Thomas~D. Cohen, Manoj K. Banerjee, Marina
Nielsen\footnote{Permanent address: Instituto de F\'\i sica,
Universidade
de S\~ao Paulo, 01498 - SP- Brazil.} and Xuemin Jin}
\centerline{\sl Department of Physics and Center for Theoretical Physics}
\centerline{\sl University of Maryland, College Park, MD \ 20742}
\vspace{7.2pt}
\centerline{March, 1994}
\vspace{14.4pt}
\begin{abstract}
Classical considerations suggest that the probability distribution
$P(R)$, where $R$ is the ratio of neutral pions to total pions emitted
from a disoriented chiral condensate (which has been hypothesized to
form in heavy ion reactions) is $R^{-1/2}/2$.  Quantum mechanical
isospin correlations between the condensate and the remainder of
the system can alter this.  Moments of the P(R) distribution  can be
expressed in terms of expectation values  of $(I^2/N^2)^m$ where $I$
is the isospin carried by the condensate, $N$  is the number of pions
emitted and m is an arbitrary integer.
We find that the probability distribution is very similar to the
classical distribution for $0.1 < R < 0.9$ unless the isospin carried
by the condensate is very large.
\end{abstract}
\newpage
%%%%%%%%%%%%%%%%%%%%%%%%%%%%%%%%%%%%%%%%%%%%%%%%%%%%%%%%%%%%%%%%%%%%%%%%
Recently, there has been considerable attention given to the possible
formation of a disoriented chiral condensate
(D$\chi$C)
in high energy collisions of  heavy ions
\cite{A,AR,BK,Bj,KT,BKT,CGM,RW,BNL}.
The basic idea is rather simple.
 During the collision a thermalized
region at high temperature may form.  If the temperature is above the
chiral restoration temperature, the chiral order parameters $\langle
\overline{q} q \rangle$ and $\langle i \overline{q} \gamma_5 \tau q
\rangle$ will be
zero in this region.  If the system subsequently cools rapidly, the
chiral-restored state becomes unstable.  The notion of  the disoriented
chiral condensate concerns the response of the system to this
instability.  Clearly the system will eventually relax to its vacuum
value in which $\langle \overline{q} q \rangle= \langle \overline{q} q
\rangle_{\rm vac} \approx - m_{\pi}^2 f_{\pi}^2 / m_q$,  and $\langle i
\overline{q} \gamma_5 \tau q \rangle$= 0 with the energy being radiated
away by hadron emission.

One  scenario for this relaxation
involves two  time scales:  a relatively quick scale in which a large
region settles into a disoriented chiral condensate and then a slower
scale in which this D$\chi$C relaxes to the physical vacuum.  The
D$\chi$C is a region which corresponds to a chirally rotated
vacuum---{\it i.e.}, a region in which the magnitude of the chiral
symmetry
breaking is the same as in the vacuum but in which the symmetry
breaking is not aligned in the $\langle \overline{q} q \rangle$ direction.
The
plausibility of this scenario stems from the following fact:  apart
from explicit chiral symmetry breaking (which is proportional to the
current quark masses and, thus,  small) and surface terms associated
with the mismatch of the chiral direction in the condensate  and the
true vacuum (which for a large spatial volume of condensate constitutes
a small fraction of the energy driving the system) all directions in
chiral space are equally likely.  The idea is that these effects which
give a preference for ``falling'' in the  $\langle \overline{q} q
\rangle$ direction are small perturbations on the scale of the
dynamical instability and hence the system is almost as likely to fall
in an arbitrary direction in chiral space.   Of course, it is clear that
if
a region of D$\chi$C forms it will eventually decay to the physical
vacuum due to the quark mass and surface effects.

Whether or not this scenario is, in fact, viable is a subject of
current interest ~\cite{RW,BNL,BK2,BVH}. The key question is whether or
not
large regions of  D$\chi$C form. Unfortunately, the detailed dynamics
of  heavy ion physics is far beyond what can be calculated directly from
QCD and
thus, studies of this kind are, of necessity based on simplified models
such as the linear sigma model~\cite{GML}.

If a large volume of D$\chi$C is formed, it will presumably decay via
the emission of hadrons with a
momentum of order $V^{-1/3}$ where $V$ is the volume of the region
with
a D$\chi$C.   Being the lightest and able to carry off chirality pions
are the most likely candidates for these hadrons.   Thus, one expects
an
anomalously large number of low momentum pions  in the rest frame of
D$\chi$C.   A second signature concerns the relative distribution of
neutral pions .

If the low momentum pions were emitted statistically, with the pions
uncorrelated with each other, then one would expect that the ratio
$R= N_0/N$, of
low momentum neutral pions to the total number of low momentum pions
should have a probability distribution given by a binomial distribution.
As $N\rightarrow \infty$ the distribution  becomes
sharply peaked at the average value, $\langle R\rangle = 1/3$ with a
variance $\langle R^2\rangle - \langle R\rangle^2 = \frac{2}{9N}$ which
goes to zero.

   It is often asserted
{}~\cite{BK,Bj,RW}, on the basis of a simple classical argument, that if
the low mass pions were to come entirely from a D$\chi C$,  then the
following nonstatistical probability distribution emerges:
\begin{equation}
P(R) = \frac{1}{2 R^{1/2}}\ .
\label{simple}
\end{equation}
The two  are so radically different that it should be easy to tell
which
 describes the pion momentum distribution better.

The derivation of eq.~(\ref{simple}) is
essentially classical \cite{A,AR}. The question we wish to address here is
whether
or not  this simple form for $P(R)$ survives a quantum treatment.
It has been pointed out previously~\cite{BKT,CGM} that if the low
momentum pions in the central rapidity region produced during the
collision from the D$\chi$C are in a  quantum state of zero isospin,
then the distribution in eq.~(\ref{simple}) will be automatically
reproduced in the large $N$ limit.  However,
  there is no guarantee that  the low momentum pions will carry
zero isospin.   Indeed, in a typical
ultrarelativistic heavy ion reaction one would expect that even if
   low momentum pions are created in large number the vast majority of
particles are ``high momentum'' pions and other energetic particles.
The total isospin of the system will be carried by both low momentum
pions   and the high momentum particles.  While in the absence of
photon emission the sum of these two isospins must equal  the total
isospin of the system, {\it a priori} there is no reason to assume that
the low momentum pions must be in  zero or any other pure isospin state.
In order to determine
how much isospin will  be carried by the low momentum pions one
requires a reliable dynamical theory of the collision process
leading to emission of low momentum pions.    Such a calculation is
well beyond the current state of the art.  Instead, we will simply
cast   the final
state  in a useful form and draw phenomenological conclusions.

Before proceeding with our calculation it is probably useful to
discuss briefly  the general question of how  the notion of a
disoriented chiral condensate can be made sensible in a  quantum
mechanical context.  One obvious point  is that the condensed state
is {\it not} the entire
quantum mechanical state.  Apart from the condensate, if  formed,
one expects a large number of  statistically emitted ``high energy''
particles  in a relativistic heavy ion reaction.  Thus,
the simplest possible description of the state would be as a product of
a  condensed state and a state describing  the rest of the system.
In fact, however, this is not viable.  In general, if the chiral
condensate
is disoriented ({\it i.e.}, not pointing in the  $\sigma$ direction in
chiral space) it points in some direction in isospin space, breaking
isospin symmetry.  We know, however, that the true quantum state does not
break isospin (if we neglect quark mass differences and electromagnetic
 effects) .  Thus, the  way to incorporate the classical notion of
the disoriented chiral condensate into a quantum picture is as follows:
the physical state, to good approximation, may be regarded as a
superposition of  product
states (between a condensed state pointing in one chiral direction and the
rest of the system).    This superposition involves a sum over states
with the condensate pointing in every  chiral direction with some
weighting function.   This notion of obtaining a quantum mechanically
 meaningful  state  from a mean field state which breaks a
symmetry
is quite standard.  For example, in nuclear structure  physics
one may encounter a Hartree-Fock solution which is deformed,
{\it i.e.}, which break rotational symmetry.  The
physical states, with good quantum numbers,  are obtained by
projection
which involves
superposing deformed states pointing in different directions.

At this point, it is convenient  to  make a few simplifications.
First,
we consider collisions of two heavy ions of zero isospin. Second, we
assume that the total isospin is conserved during the collision,
i.e., we ignore radiative processes. Third, we  assume that in such
collisions   pions are produced with essentially zero momentum in the
rest frame of the condensate. This is
possible only if the region occupied by these pions is very large.
We consider  events in which the number of zero momentum
pions is large assuming that such
events occur during  heavy ion collisions.

The concept of a disoriented chiral condensate aligned in a nontrivial
isospin direction
is meaningful only if essentially all of the pions in the condensate
will be aligned in the
same isospin direction in the condensate.  This trivially implies
that
the pions in the condensate are in a completely symmetric state in
isospin and by Bose symmetry they must be in a symmetric spatial
state.
All of the physical consequences discussed in this letter depend
essentially on the low momentum pions being in an isospin symmetric
state.  We should note that  the
signatures discussed in this letter can tell us  only  whether
a state of low momentum pions is  symmetric in isospin.  We have
no way to determine whether such a state has its origin in
a disoriented chiral condensate or via some other (as yet unknown)
mechanism.    We will refer to any state containing only low momentum
pions in an isospin symmetric state as a condensate
regardless of the actual mechanism by which it is formed.

We will study moments of the probability distribution because it is
possible to obtain compact analytical expressions for these quantities
in the limit of large $N$. Calculation of $P(R)$ as a function of $R$
involves differences of large numbers and becomes rather unwieldy for
$N\,>\,100$. We will show only a few illustrative results for $N=100$.

To begin, we group the particles in the final
state into two classes, low momentum pions and everything else.
Assuming
that the low momentum pions are all zero momentum we see that one
completely labels the state by three quantum numbers which can be
either the numbers of $\pi^\pm$ and $\pi^0$,  or,   the total number
of
pions $N$, the total isospin $I_c$, and the third component of the
isospin $i_3$. We write the
most general final state with {\it zero isospin} in the form:
\begin{equation}
|\psi; I=0 \rangle = \sum_{N,I_c, i_3}
C_{N,I_c}\frac{(-)^{i_3}}{\sqrt{2I_c+1}}  |N,I_c,i_3 \rangle_{\rm low}
 |\psi_{N}; I_c, -i_3 \rangle_{\rm high},
\label{simon}
\end{equation}
where $C_{N,I_c}$ is a numerical coefficient, the subscript ``low''
labels the states of the condensate containing low momentum  pions.
The notation $|\psi_{N}; I_c, -i_3 \rangle_{\rm high}$ denotes the
state containing no low momentum pions and carrying isospin $I_c$
and  third component of isospin
$-i_3$; the subscript $N$ merely indicates that it is the state
coupling
to a state of $N$ low momentum pions. If all states are
normalized one has the constraint that
\begin{equation}
\sum_{N,I_c} |C_{N,I_c}|^2 =1.
\label{Norm}
\end{equation}
Quantum mechanical correlations
between the isospin carried by the
low momentum pions and the rest of the system are exhibited by
eq.~(\ref{simon}).

If all the low momentum pions are zero momentum,
the low momentum pion states $| N, I_c, i_3 \rangle_{\rm low}$ may
be generated from vacuum by application of
the  creation operators for the three types of pions.
Here, it is convenient to work in a cartesian basis for the isospin
and
thus the creation operators are $a^{\dagger}_i$ with $i=1, 2, 3$.
Given the assumption of  only zero momentum pions, the normalized low
momentum states of fixed $N$,  $I_c$ and $i_3$ are unique
\begin{equation}
| N, I_c, i_3 \rangle_{\rm low} = \int \, {\rm d}\Omega \,  {\rm
Y}_{I_c, i_3}(\Omega) \, (N !)^{-1/2} (\hat{n}(\Omega) \cdot a^{\dagger} )
^N |vac \rangle
\label{lowstate}
\end{equation}
where $\hat{n}(\Omega) $ is a unit vector in the $\Omega$ direction and
${\rm Y}_{I_c, i_3}(\Omega)$ is the spherical harmonic function.

Consider an operator, ${\cal O}_{\rm low}$ which acts only in the
space of the low momentum pions, i.e., an operator whose matrix
elements do not depend on  $|\psi_{N};I_c,-i_3\rangle_{\rm high}$ at
all.   Since, the operator is unity in the ``high'' space, the
expectation value of such an operator is simple:
\begin{equation}
 \langle{\cal O}_{\rm low}\rangle =\sum_{N,I_c}| C_{N,I_c}
|^2\widetilde{{\cal O}_{\rm low}}\ ,
\label {simpleme}
\end{equation}
where
\begin{eqnarray}
 \widetilde{{\cal O}}_{\rm low}(I_c)&\equiv &\sum_{i_3} {1\over 2I_c+1}
\langle N,I_c,i_3 | {\cal O}_{\rm low}| N,I_c,i_3 \rangle_{\rm low}
\label {reduceme} \\
&=&\sum_{i_3} {1\over 2I_c+1} \langle N,I_c,i_3 | {\cal O}_{\rm low}
^{\rm isoscalar\,\, part} | N,I_c,i_3 \rangle_{\rm low} \nonumber\; .
\end{eqnarray}
Note that the number of low momentum pions $N$, the number
of low momentum neutral pions $N_0$, and the ratio $R= N_0/N$ are
all examples of operators of the form of  ${\cal O}_{\rm low}$. For
${\cal O}_{\rm low}=R^m$,
the problem of calculating the moments of the distribution is reduced
to
evaluating matrix elements  $\widetilde{R^m}$. The  general result is
given below:
\begin{equation}
\lim_{N\rightarrow\infty}\widetilde{R^m}=\sum_{i=0}^{[m/2]}\sum_{j=0}^
{i}\left\{{(-1)^{i+j} m!\over 2^{2i}  i! j! (m-2i)! (i-j)! [2j+2(m-i)+1]}
 \left({I_c(I_c+1)\over N^{2}}\right)^i\right\}\; .
\label{moment-q}
\end{equation}
It is straightforward to obtain expressions for $\langle R^m\rangle$ in
the limit $N\rightarrow\infty$  by using eqs.~(\ref{simpleme})
and (\ref{moment-q}) and the obvious definition, $\langle I^2/N^2
\rangle = \sum_{N,I_c}|C_{N,I_c}|^2 I_c(I_c+1)/N^2$. Writing
\begin{equation}
\lim_{N\rightarrow\infty}
\langle R^m\rangle=\sum_{i=0}^{[m/2]}b^m_i\left\langle {I^{2i}
\over N^{2i}}\right\rangle \ ,
\label{fi-mom}
\end{equation}
the results for the coefficients $b^m_i$ for $m=1$ through $6$ are
given in Table~\ref{moment-tab}.
We note that these results agree with the classical result   when
$\langle\frac{I^2}{N^2}\rangle \ll 1$.

\begin{table}{}
\begin{center}
\begin{tabular}{|l|l|l|l|l|}\hline\hline
Moment & $1$ & $\langle\frac{I^2}{N^2}\rangle $ & $\langle
\frac{ I^4}{N^4}\rangle $& $\langle\frac{I^6}{N^6}\rangle $ \\
\hline
$\langle R \rangle $&$ \frac13 $& & & \\ \hline
$\langle R^2 \rangle $&$ \frac15 $&$ -\frac{1}{15} $& & \\ \hline
$\langle R^3 \rangle $&$ \frac17 $&$ -\frac{3}{35} $& & \\ \hline
$\langle R^4 \rangle $&$ \frac19 $&$ -\frac{2}{21} $&$\frac{1}{105}$&
\\ \hline
$\langle R^5 \rangle $&$ \frac{1}{11} $&$ -\frac{10}{99} $&$
\frac{5}{231} $& \\ \hline
$\langle R^6 \rangle $&$ \frac{1}{13} $&$ -\frac{15}{143} $&$
\frac{5}{143} $&$ -\frac{5}{3003}$  \\ \hline
\end{tabular}
\end{center}
\caption{ The   coefficients  $b^m_i$ defined in
eq.~(\protect{\ref{fi-mom}})
 .}
\label{moment-tab}
\end{table}

  One sees from Table~\ref{moment-tab} that when $\langle\frac{I^2}
{N^2}\rangle $ increases from $0$ to $0.1$ that moments of order $4$
and higher   change by as much as $\sim 10$\%. Thus an experimental
 determination of high moments with sufficiently good accuracy can
provide a measure of the isospin content of the condensate.

Fortunately,   there is a much more promising
experimental way of relating $\langle\frac{I^2}{N^2}\rangle $ to
observable quantities.  It has been noted by Greiner,
Gong and M\"{u}ller~\cite{CGM} that, if the low momentum  pions come
from an $I_c
= 0$ condensate then in {\it every} event the number of low momentum
$\pi^+$'s must exactly equal the number of $\pi^-$'s. Conversely,
difference between the two numbers provides a direct measure of the
isospin content: \begin{equation}
\lim_{N\rightarrow\infty}\left<{(N_+ - N_-)^{2i}\over N^{2i}} \right>
= {1\over 2i+1}\left<{I^{2i}\over N^{2i}}\right>\ .
\label{NpNm}
\end{equation}
It is easy to derive the relation by noting that $N_+ - N_- = i_3$
and using eqs.~(\ref{simpleme}) and (\ref{reduceme}). Thus a
measurement of $\langle{(N_+ - N_-)^2\over N^2} \rangle$ is the most
direct and natural way to measure the isospin content of the condensate.
In particular, the variance is given by
\begin{equation}
\langle R^2 \rangle - \langle R \rangle^2  ={4\over 45} -
\left<{(N_+ - N_-)^2\over 5N^2}\right>\ .
\label{variance-final}
\end{equation}

The result in eq.~(\ref{variance-final}) is in a real sense model
independent. It does, however, depend rather strongly on two
underlying assumptions.
 The first is the assumption that a single large region of D$\chi$C
is
formed.  If multiple regions form with different orientations the
results will certainly change, and the variance in the $R$
distribution will certainly be reduced.   One might hope to use
eq.~(\ref{variance-final}) to experimentally test whether a single
region of
D$\chi$C is, in fact produced. The second important assumption is that
the total system (or at least the part of the system which ultimately
contributes) is  isospin zero.  This will certainly be violated to a
certain degree.  The initial state may well not be isospin zero.  More
importantly,  photons
will be emitted during the heavy ion reaction which violate isospin
conservation.  It is essential to estimate how large an effect these
isospin violating effects will have on the final results.  Unfortunately,
estimates of the corrections to  eq.~(\ref{variance-final}) due to
isospin
violation will, of necessity be model dependent.

We have discussed the effect of quantum mechanical isospin correlations
on moments of the $P(R)$ distribution.
It should be remembered, of course, that the moments are derived
quantities---the primary quanity is $P(R)$ itself.  We would like
to understand the effect  of  isospin correlations on $P(R)$.
We write
the
probability distribution  as
\begin{equation}
P(R)=\sum_{N,I_c}|C_{N,I_c}|^2 P_{I_c}(R)\ ,
\label{P}
\end{equation}
where the probability of finding $N_0$
neutral pions in the state $|N,I_c,i_3 \rangle_{\rm low}$
is given by
\begin{equation}
P_{I_c}(R)={1\over 2I_c+1}\sum_{i_3}  |\langle N,I_c,i_3 |
N,I_c,N_0 \rangle_{\rm low}|^2 \; .
\end{equation}
Following  Horn and Silver \cite{HS}, this probability is evaluated
to be
\begin{eqnarray}
P_{I_c}(R)={1\over 2I_c+1}\sum_{i_3} N_0!\left({N-N_0+i_3
\over 2}
\right)!\left({N-N_0-i_3\over 2}\right)!{(2I_c+1)!!(N-I_c)!!
\over
(N+I_c+1)!!}&\times&\nonumber\\
{2^{I_c-i_3-N_0}\over\sum_j{2^{-2j}\over(i_3+j)!j!
(I_c-i_3-2j)!}}\left(\sum_k{(-1)^k2^{-2k}\over
k!(I_c+k)!(I_c-i_3-2k)!\left({N-N_0-i_3
-2k\over 2}\right)!\left({N_0+i_3-I_c+2k\over 2}\right)!}\right)^2 &,&
\label{PIc}
\end{eqnarray}
where the sums are over all integers $i_3$, $j$ and $k$ such that the
factorials can be defined.

We see clearly that, the  quantum mechanical isospin correlations will
also affect the probability distribution. Needless to say, we cannot
calculate the distribution without a knowledge of the coefficients
$C_{N,I_c}$, which, in turn, requires a knowledge of the reaction
mechanism. However, we can learn something by examining   $P_{I_c}(R)$
for a range of values of $I_c$. These probability distributions have
sharp odd-even variations. Such variations will never be seen in an
experiment where one measures $P(R)$. The sum over $I_c$ in
eq.~(\ref{P}) smoothes out the variations. In addition, there is
bound to be additional averaging over neighboring $N_0$ and $N$
because of the difficulty of counting the pions exactly.   We
suppress these oscillations partially by averaging over two neighboring
numbers of neutral pions.  We believe that the procedure improves
markedly the appearance of the graphs without affecting the conclusions
we draw. In Fig.1 we show these partially averaged $P_{I_c}(R)$ vs. $R$
for $I_c=0$, $10$, $20$, $30$ and $60$ for $N=100$. The distributions
for $I_c=10$ and $20$ are not distinguishable from the $I_c=0$
distribution which, one may recall, agrees with the classical
distribution~\cite{ftnt1}. The distribution for $I_c=30$ is
barely distinguishable. These graphs tell us that the classical
result of eq.(\ref{simple}) is  {\it robust} (except at the endpoint
regions near $R=0$ and $R=1$) and will describe the actual distribution
$ P(R)$ unless there is massive isospin mixing. Conversely, agreement
with the classical result will tell us  little about the isospin
content of the condensate. It will establish that the pions in the
condensate are in fully isosymmetric state and, hence, in a fully
space symmetric state.  Information about the isospin admixtures is
best obtained by studying the charge asymmetry distribution.

T.D.C. thanks Berndt  M\"{u}ller for introducing him to the problem
and
for helpful discussions.  T.D.C., M.K.B., and X.J. thank the US
Department of
Energy for support of this research through grant No. DE-FG02-93ER-40762.
 T.D.C. and X.J. also
acknowledge the support of the US National Science Foundations
Presidential Young Investigator program through grant No. PHY-9058487.
M.N. acknowledges the warm hospitality and congenial atmosphere provided
by the Nuclear Theory Group of the University of Maryland and support
from FAPESP-Brazil .
\eject

\eject
\noindent
{\Large\bf Figure Captions}\\ \\ \\
{\bf Figure 1.} $P_{I_c}(R)$  vs. $R$ for $I_c=0$, $10$, $20$, $30$ and
$60$ for N=100. The graphs for $I_c=10$ and $20$ are not distinguishable
from that for $I_c=0$ and are not labeled. Other graphs are labeled by
their isospins.

\begin{thebibliography}{100}
\bibitem{ftnt1} To be precise $P_{\rm classical}=\frac12 P_{I_c=0}(R)$
for even $N_0$ only, the latter being zero for odd $N_0$. Here the
factor of $\frac12$ gets included through the averaging over neighboring
odd and even $N_0$s.
\end{thebibliography}
\end{document}